\documentclass[msmath,amssymb,aps,pra,twocolumn,epsfig,showpacs,bibliography,
lengthcheck,superscriptaddress]{revtex4-1}

\usepackage{graphicx}
\usepackage{dcolumn}
\usepackage{bm}
\usepackage{hyperref}
\usepackage{epstopdf}
\usepackage{subcaption}
\usepackage{caption}
\usepackage{footmisc}
\usepackage{amsmath}

\begin{document}
\title{Herriott-cavity-assisted all-optical atomic vector magnetometer}
\author{B. Cai}
\author{C.-P. Hao}
\affiliation{Hefei National Laboratory of Physical Sciences at the Microscale, University of Science and Technology of China, Hefei 230026, China}
\affiliation{Department of Precision Machinery and Precision Instrumentation, Key Laboratory of Precision Scientific Instrumentation of Anhui Higher Education Institutes, University of Science and Technology of China, Hefei 230027, China}

\author{Z.-R. Qiu}
\affiliation{Hefei National Laboratory of Physical Sciences at the Microscale, University of Science and Technology of China, Hefei 230026, China}
\author{Q.-Q. Yu}
\affiliation{Hefei National Laboratory of Physical Sciences at the Microscale, University of Science and Technology of China, Hefei 230026, China}
\affiliation{Department of Precision Machinery and Precision Instrumentation, Key Laboratory of Precision Scientific Instrumentation of Anhui Higher Education Institutes, University of Science and Technology of China, Hefei 230027, China}
\author{W. Xiao}
\affiliation{State Key Laboratory of Advanced Optical Communication Systems and Networks, School of Electronics Engineering and Computer Science, and Center for Quantum Information Technology, Peking University, Beijing 100871, China}

\author{D. Sheng}
\email{dsheng@ustc.edu.cn}
\affiliation{Hefei National Laboratory of Physical Sciences at the Microscale, University of Science and Technology of China, Hefei 230026, China}
\affiliation{Department of Precision Machinery and Precision Instrumentation, Key Laboratory of Precision Scientific Instrumentation of Anhui Higher Education Institutes, University of Science and Technology of China, Hefei 230027, China}

\begin{abstract}
We report an all-optical atomic vector magnetometer using dual Bell-Bloom optical pumping beams in a Rb vapor cell. This vector magnetometer consists of two orthogonal optical pumping beams, with amplitude modulations at $^{85}$Rb and $^{87}$Rb Larmor frequencies respectively. We  simultaneously detect atomic signals excited by these two pumping beams using a single probe beam in the third direction, and extract the field orientation information using the phase delays between the modulated atomic signals and the driving beams. By adding a Herriott cavity inside the vapor cell, we improve the magnetometer sensitivity. We study the performance of this vector magnetometer in a magnetic field ranging from 100~mG to 500~mG, and demonstrate a field angle sensitivity better than 10~${\mu}$rad/Hz$^{1/2}$ above 10~Hz.
\end{abstract}

\maketitle
\section{Introduction}
Optically pumped atomic magnetometers use polarized spin to detect the external field~\cite{budker2007}, and have demonstrated wide applications in research fields as fundamental physics~\cite{griffith2009,pospelov2013}, material science~\cite{knappe2010,maser2011,hu2019}, geophysics~\cite{prouty2013} and biomagnetic imaging~\cite{kim2014,alem2015,boto18}. Recently there have been two main development directions in atomic magnetometry. One is to improve the sensor sensitivity, which has reached the subfemtotesla level~\cite{Lee2006,dang2010}. The other one is miniaturization for smaller size and power consumption~\cite{liew2004,mhaskar12}. It is common to sacrifice the sensor performance for miniaturization~\cite{kitching2018}, however, this problem can be avoided by introducing multipass cavities~\cite{li2011,jensen2014,clevenson2015}. The cavity-assisted atomic magnetometers have been used to enhance the detection of nuclear magnetic and quadrupole resonance signals~\cite{shi2013,cooper2016}, improve the field sensitivity to the standard quantum limit level~\cite{sheng2013}, and realize magnetoencephalogram detection in unshielded environment~\cite{limes2020}.

In addition to the magnetic field magnitude, it often requires the whole field vector information in precision measurements~\cite{afach2015} and geophysics applications~\cite{prouty2013}. Currently most atomic vector magnetometers are based on modified scalar magnetometers. The field orientation has been extracted using methods such as effective modulations in each axis ~\cite{patton2014,huang2015}, absorptions of multiple beams~\cite{afach2015}, or combined multiple harmonics from the Faraday rotation signals~\cite{huang2016,ingleby2018,pyragius2019}.  Many of these methods involve additional rf fields.  It is preferred to operate atomic magnetometers purely using light-atom interactions in practice~\cite{prouty2013}, because it is easier to calibrate the direction and control the interaction regions of laser beams compared with rf fields. This all-optical vector magnetometer operation has been demonstrated for several different schemes in a bias field of 10~mG level ~\cite{patton2014,huang2015,sun2017}. In this paper, we report a compact, high bandwidth, and highly sensitive all-optical vector magnetometer using the Bell-Bloom optical pumping methods~\cite{bell1961}. We improve the magnetometer sensitivity by implementing a Herriott cavity~\cite{silver2005} to the vapor cell. In a spherical coordinate, the magnetic field vector is described by its magnitude and two orientation angles. While the field magnitude is read from the magnetometer operating in the scalar mode, the field orientation is extracted from the phase delays between the modulated atomic signals and the driving beams. We study the performance of a vector magnetometer working with this method in a field magnitude ranging from 100~mG to 500~mG, and demonstrate a field angle sensitivity better than 10~${\mu}$rad/Hz$^{1/2}$ above 10~Hz.

\section{Experiment Methods}
\begin{figure}[htb]
\includegraphics[width=3in]{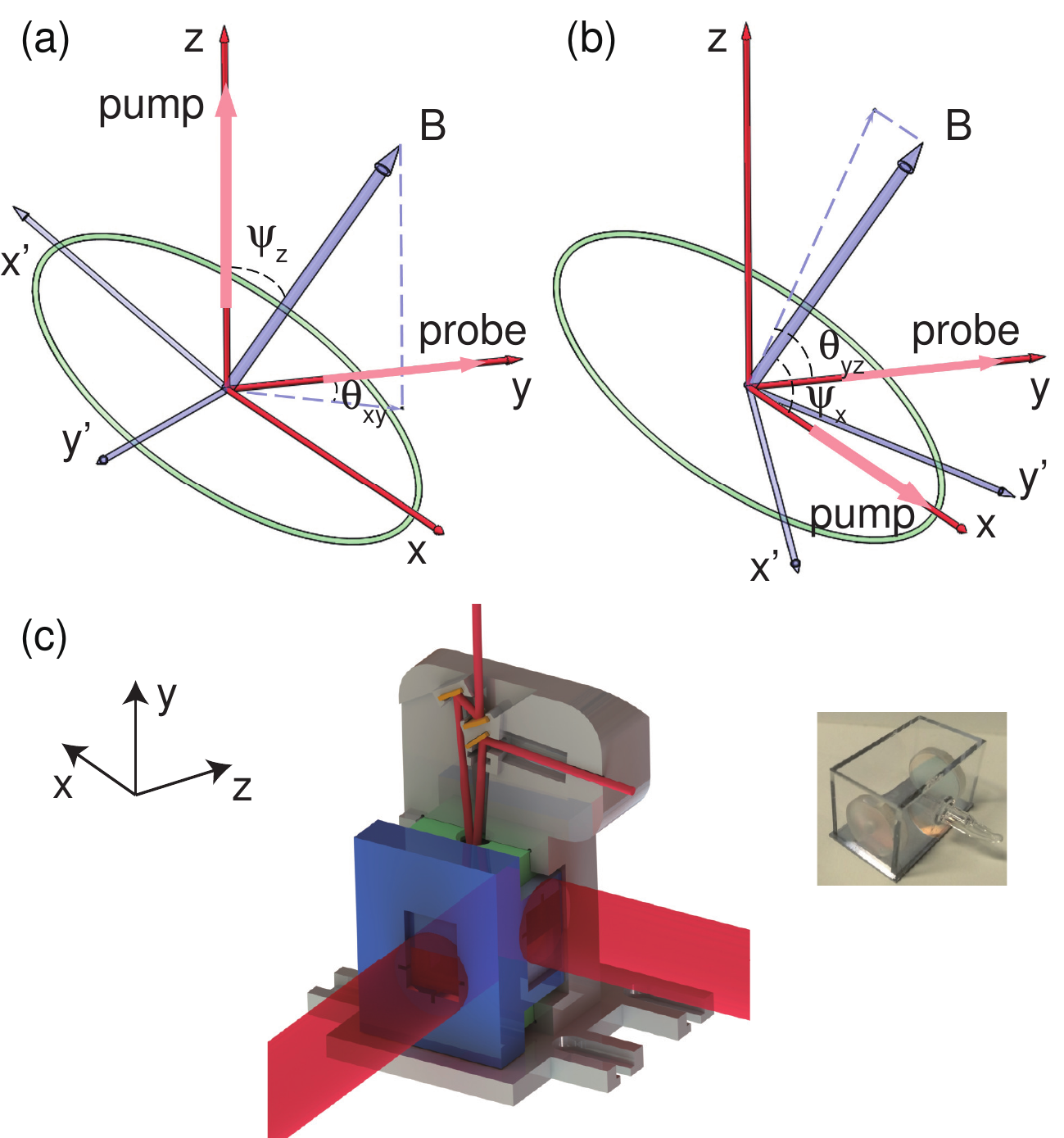}
\caption{\label{fig:setup}(Color online) Plots (a) and (b) show two configurations of Bell-Bloom magnetometers with pumping beams sent from $z$ and $x$ axes, respectively. Plot (c) shows the experiment setup, with a picture of the atomic vapor cell used shown on the right.}
\end{figure}

Figure~\ref{fig:setup}(a) shows a Bell-Bloom magnetometer configuration with an arbitrary magnetic field direction. With the pumping beam intensity modulated at a frequency $\omega$, the optical pumping rate is $R_{op}=a_0+\sum_n{a_n}\cos(n\omega{t}-\alpha_n)$. When $\omega$ is close to the atomic Larmor precession frequency $\omega_L$, the atomic polarization $P$ is most efficiently pumped in the transversal plane perpendicular to the magnetic field, denoted as the $x'y'$ plane in Fig.~\ref{fig:setup}(a). When $R_{op}\ll\omega$, the component of atomic polarization modulated at the frequency $\omega$ is mainly in the $x'y'$ plane, with the expression as~\cite{bell1961}
\begin{equation}
\label{eq:bb}
P_{x'}+iP_{y'}=\frac{a_1\sin\psi_z}{2\left[R+i(\omega-\omega_L)\right]}e^{-i(\omega{t}-\alpha_1)},
\end{equation}
where $R=a_0+R_d$, and $R_d$ is the depolarization rate in the absence of optical pumping. In cases of near resonance, $|\omega-\omega_L|\ll{R}$, the projection of the rotating atomic polarization in Eq.~\eqref{eq:bb} along the probe beam direction is:
\begin{equation}
\label{eq:py}
P_y=\frac{a_1\sin\psi_z\sqrt{\cos^2\psi_z\cos^2\theta_{xy}+\sin^2\theta_{xy}}}{2R}\sin(\omega{t}+\phi_1).
\end{equation}
Since the probe signal is proportional to $P_y$, $\phi_1$ in Eq.~\eqref{eq:py} is the phase delay between the modulated atomic signal and the driving beam, with the expression as
\begin{equation}
\label{eq:phi}
\phi_1=\cot^{-1}(\cos\psi_z\cot{\theta_{xy}})+\beta,
\end{equation}
where $\beta=\pi/2-\alpha_1+(\omega-\omega_L)/R$.

We get the amplitude and phase information of a Bell-Bloom magnetometer simultaneously from the phase-sensitive detection of the probe signal. While the signal amplitude is used to track the field magnitude according to Eq.~\eqref{eq:bb}, one degree of freedom of the magnetic field orientation can be fixed by measuring the relative phase $\phi_1$ in Eq.~\eqref{eq:phi}. Another independent equation is required to fully reconstruct the magnetic vector, which is realized by adding a second pumping beam in the $x$ direction as shown in Fig.~\ref{fig:setup}(b). To separate the contributions of the two pumping beams in the probe signal, we fill atoms of both Rb isotopes in the vapor cell, and selectively modulate the pumping beam intensities at the Larmor frequencies of corresponding isotopes. Using this scheme, the second independent equation is
\begin{eqnarray}
\phi_2&=&\cot^{-1}(-\cos\psi_x\cot{\theta_{yz}})+\beta\label{eq:phi21}\\
&=&\cot^{-1}(-\frac{1}{2}\sin\psi_z\tan\psi_z\sin{2\theta_{xy}})+\beta\label{eq:phi2}.
\end{eqnarray}
$\theta$ and $\psi$ are directly extracted from the combined Eqs.~\eqref{eq:phi} and ~\eqref{eq:phi2}. It is also convenient to construct a physical picture using a different combination of Eqs.~\eqref{eq:phi} and ~\eqref{eq:phi21}, where each equation defines a curve on a unit sphere, and the magnetic field direction points to the cross point of the two curves.

In the experiment, we fabricate an atomic vapor cell with Rb atoms in natural abundance, 400~torr N$_2$ gas, and an anodic bonded Herriott cavity inside~\cite{liew2004,cooper2016}. This Herriott cavity consists of two cylindrical mirrors with a curvature of 100~mm, a diameter of 12.7~mm, a thickness of 2.5~mm, a relative angle between symmetrical axes of 52$^\circ$, and a separation of 18.9~mm. A linearly polarized probe beam, 30~GHz blue detuned from the Rb D$_1$ transition with a diameter of 1 mm, enters the cavity from a hole in the center of the front mirror, and exits from the same hole after 21 times of reflections. The Faraday rotation of the probe beam is analyzed by a polarimeter. As shown in Fig.~\ref{fig:setup}(c), a three-dimensional printed platform, containing the probe beam optics and a vapor cell with inner dimensions of 14.5$\times$27$\times$15~mm$^3$, sits in the center of five-layer magnetic shields. Two  distributed-Bragg-reflector laser diodes provide the pumping and probe beams. A pumping beam on resonance with the pressure broadened Rb D$_1$ line is split into two parts, which pass through the vapor cell from the $x$ and $z$ directions, respectively,  with each beam diameter of 1~cm and a separation of beam centers of 9~mm in the $y$ direction. As mentioned previously, these two beams are independently amplitude modulated at the Larmor frequencies of $^{85}$Rb ($z$ axis beam) and $^{87}$Rb ($x$ axis beam) using acoustic-optical modulators (AOMs). We reduce the cross-talk between the two beams by spatially separating their interaction regions with atoms using a printed mask, which guides the pumping beams through two square apertures with no intersections in the $y$ axis as shown in Fig.~\ref{fig:setup}(c).

We lock the laser beam powers by sending feedback signals the amplitude control of AOMs. The power of the output probe beam from the Herriott cavity is locked at 0.7~mW by using signals from polarimeter channels filtered by low pass filters as the feedback signals. Each pumping beam intensity is monitored by splitting a part of its beam into two photodiode detectors before entering the cell. One monitor signal is used for pumping beam power locking, and the other one is used as the reference signal for demodulating the magnetometer signals. The phase delay  results between the atomic and reference signals are recorded when the pumping beam modulation frequencies are on resonance. Due to the pressure broadening of atomic transitions, we did not add feedback to laser frequencies.


\section{Experiment Results}
\begin{figure}[htb]
\includegraphics[width=3in]{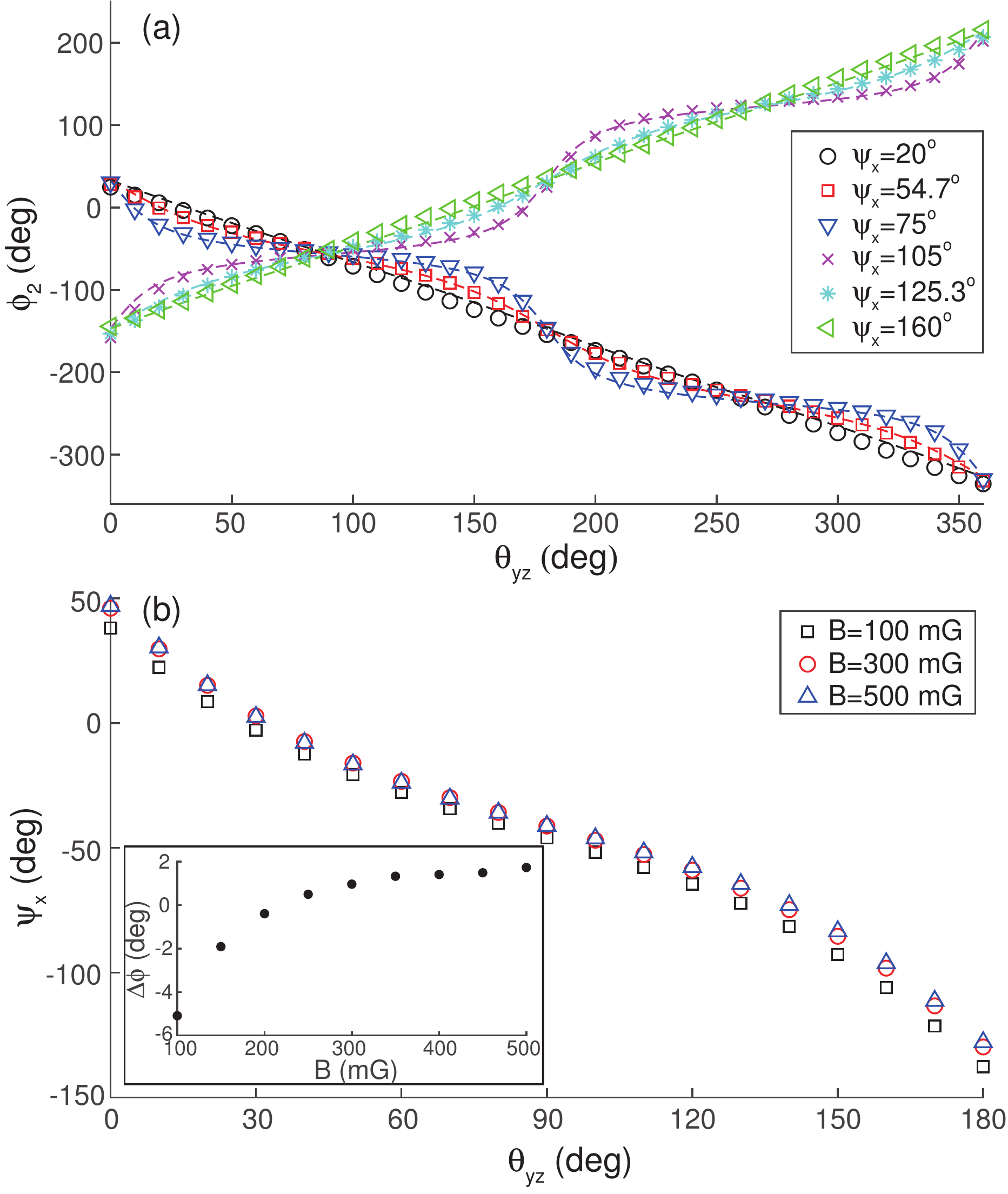}
\caption{\label{fig:phasetheta}(Color online) (a) Comparison of experiment data (symbols) and theoretical calculations using Eq.~\eqref{eq:phi21} (dash lines) of the phase delay results. (b) Experimental data of $\phi_2$ for different B fields at $\psi_x=54.7^\circ$, and the inset shows the average phase differences between the lines in plot (b). }
\end{figure}

We use three home-built current sources to control the magnetic field vector inside the magnetic shields. Figure~\ref{fig:phasetheta}(a) shows the phase results demodulated from the $^{87}$Rb signals for different field orientations with a bias field magnitude of 100~mG, cell temperature of 85 $^\circ$C, each pumping beam intensity of 25~mW/cm$^2$, and modulation duty cycle of 20\%. By adding a constant in Eq.~\eqref{eq:phi21} to account for extra phase shifts in the experiment, we plot theoretical calculations and experiment data in the same figure. We find the experimental data agrees with the theoretical values within $\pm5^\circ$ for the studied field angle range covering [$\psi_x=20^\circ-75^\circ$, $\theta_{yz}=0^\circ-360^\circ$] and [$\psi_x=105^\circ-160^\circ$, $\theta_{yz}=0^\circ-360^\circ$]. If we leave the constant added in Eq.~\eqref{eq:phi21} as a free parameter for different $\psi_x$ curves, the discrepancies between the experimental data and theoretical values decrease to $\pm2^\circ$. We are investigating this discrepancy for better accuracy of the theoretical model.

As a general property of atoms driven by resonant fields, the phase delay results of this vector magnetometer are independent of the magnetic field magnitude. Figure~\ref{fig:phasetheta}(b) shows the experimental data of $\phi_2$ for several different magnetic fields with the same zenith angle $\psi_x=54.7^\circ$. While the shape of $\phi_2$ remains the same, there is a shift of $\phi_2$ as field magnitude $B$ changes. The average value of the phase shift is shown in the inset of Fig.~\ref{fig:phasetheta}(b). We find from the experimental data that this shift is independent of the pumping beam power, and also find from density matrix simulations~\cite{hao2019} that the effect of spin-exchange interactions is negligible in the experiment conditions. We attribute this shift mainly to the changes of the Bell-Bloom magnetometer line shape as $B$ increases, which leads to a systematic error of the fitted resonance used in the experiment, and introduces a corresponding shift of $\beta$ in Eq.~\eqref{eq:phi2}.

\begin{figure}[htb]
\includegraphics[width=3in]{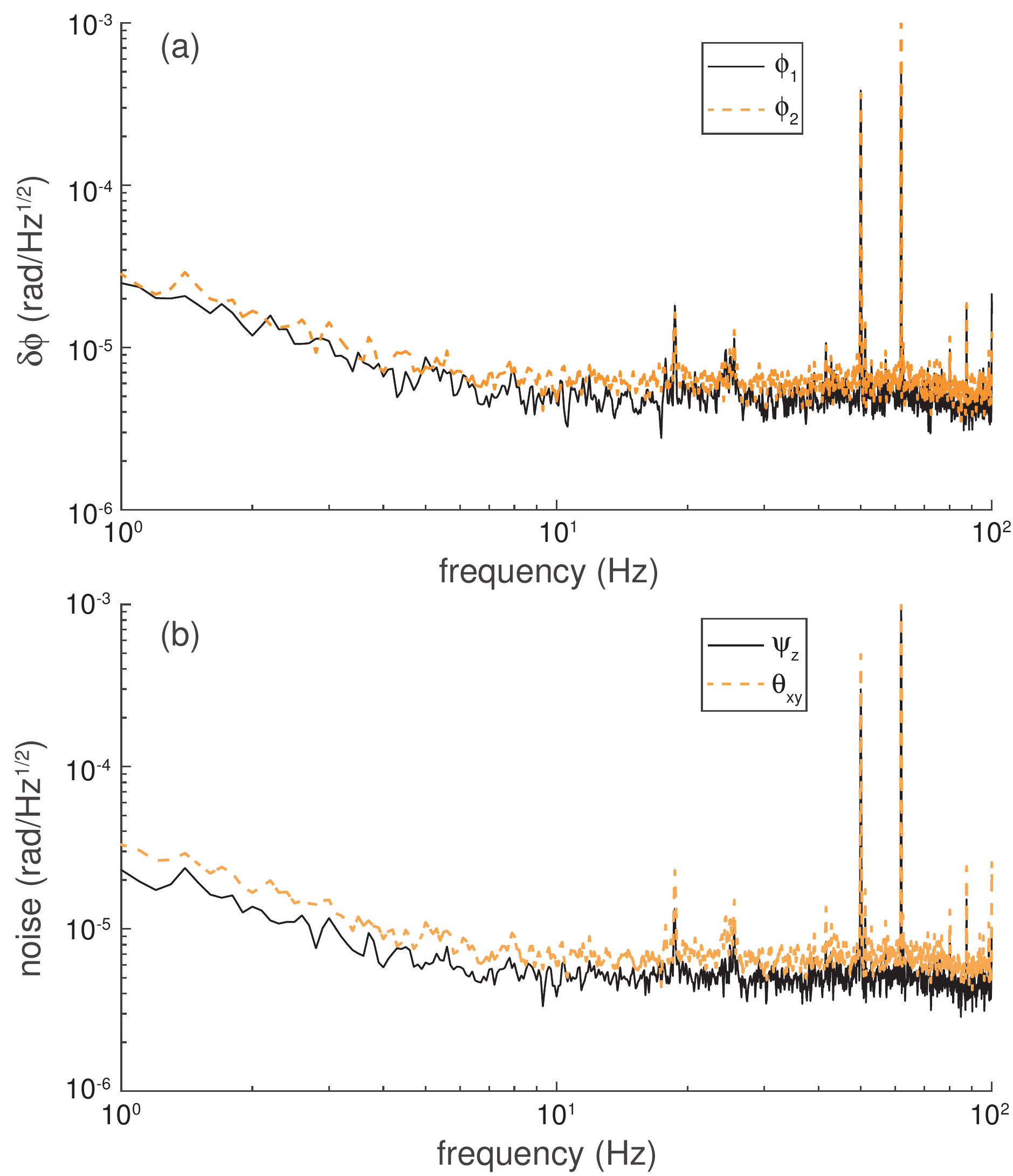}
\caption{\label{fig:noise}(Color online) (a) Phase result noise and (b) field angle noise of the vector magnetometer with the experimental conditions in the main text.}
\end{figure}

The vector magnetometer studied in this paper shares the same magnetometer bandwidth with the normal Bell-Bloom magneotmeter, which is determined by the pumping beam power broadening. In the experiment, we limit the bandwidth to 100~Hz using the lock-in amplifier filters. According to Eq.~\eqref{eq:phi}, the phase result noise $\delta\phi$ of the vector magnetometer is proportional to the magnetic field magnitude noise $\delta{B}$ as
\begin{equation}
\label{eq:noise}
\delta{\phi}=\frac{\gamma\delta{B}}{R},
\end{equation}
where $\gamma$ is the atomic gyromagnetic ratio.  With the field vector as $B=100~$mG, $\psi_z=54.7^\circ$, $\theta_{xy}=45^\circ$, and same pumping beam parameters as Fig.~\ref{fig:phasetheta}(a), we first measured the scalar Bell-Bloom magnetometer signals using a method similar to Ref.~\cite{smullin2009}, and found a field magnitude sensitivity of 0.5~pT/Hz$^{1/2}$ (0.7~pT/Hz$^{1/2}$) above 10~Hz for $^{85}$Rb ($^{87}$Rb) atoms. While the magnetometer was operated in the vector mode, we directly measured the phase result noises of $\phi_1$ and $\phi_2$  from the demodulated signals as shown Fig.~\ref{fig:noise}(a), which are also consistent with the calculations of Eq.~\eqref{eq:noise} using the scalar mode measurement results and calculated $R=2100$~s$^{-1}$ (3200~s$^{-1}$) for $^{85}$Rb ($^{87}$Rb) atoms considering the nuclear slowing factors~\cite{appelt98,smullin2009}. It is straightforward to show that, when $\theta_{xy}$ is close to $\pi/4$, the field angle noises are related to the phase result noise as
\begin{eqnarray}
\label{eq:anglenoise}
\delta\theta_{xy}&=&\frac{1}{2}\sqrt{(\tan\psi_z\delta\psi_z)^2+[(\cos\psi_z+\sec\psi_z)\delta\phi_1]^2},\nonumber\\
\delta\psi_z&=&\frac{4+\sin^2\psi_z\tan^2\psi_z}{(4+2\tan^2\psi_z)\sin\psi_z}\delta\phi_2.
\end{eqnarray}
We plot field angle noises using Eq.~\eqref{eq:anglenoise} in Fig.~\ref{fig:noise}(b), which demonstrates a vector magnetometer sensitivity better than 10~${\mu}$rad/Hz$^{1/2}$ above 10~Hz for both azimuthal and zenith angles. This sensitivity is mainly limited by the noises of the magnetic field current sources.

\begin{figure}[htb]
\includegraphics[width=3in]{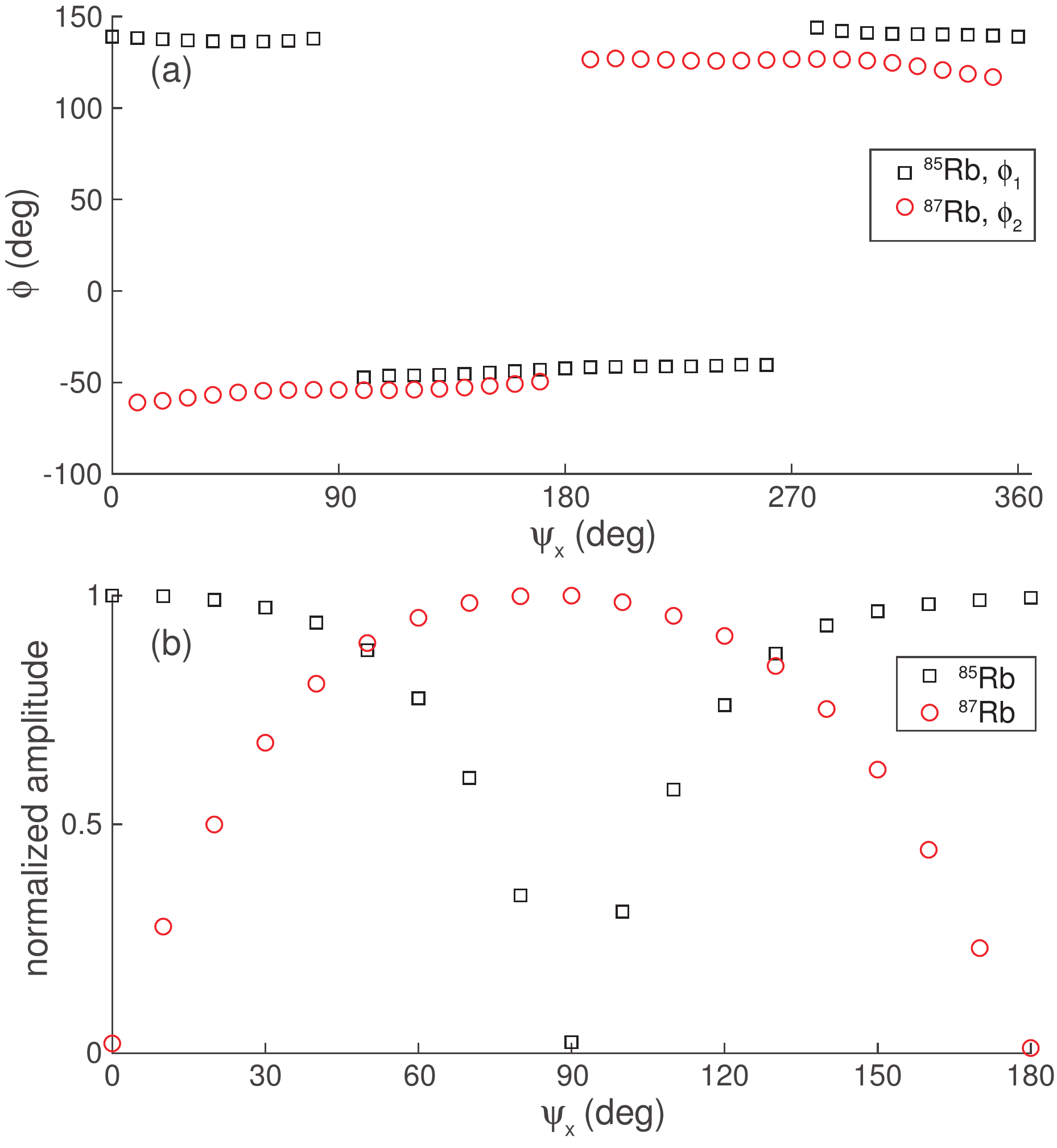}
\caption{\label{fig:deadzone}(Color online) (a) Demodulated phase and (b) amplitude results of the vector magnetometer signals when the field orientation is near the $xz$ plane.}
\end{figure}

There are several regions of field orientations where at least one of the field angle noises is significantly larger than the noise level in Fig.~\ref{fig:noise}(b) using the phase readout method. According to Eqs.~\eqref{eq:phi} and ~\eqref{eq:phi2}, these regions are around the field angles of $\psi_z=\pi/2$ or $\theta_{xy}=n\pi/2$. As an example of the dead zone, we plot the phase results for magnetic field orientations near the $xz$ plane in Fig.~\ref{fig:deadzone}(a), where the weak relation between $\phi$ and $\psi_x$ leads to a large noise on the determination of the zenith angle. However, the corresponding signal amplitude shows a strong dependence on $\psi_x$ as shown in Fig.~\ref{fig:deadzone}(b), and we can eliminate this dead zone by combining both the signal phase and amplitude information. Following this approach, we can largely reduce the vector magnetometer dead zones to a small region of field orientations near the probe beam direction, where the signal amplitude is limited by the small projection of atomic polarization in the probe beam direction. This dead zone could be eliminated by adding another probe beam in a different direction.

\section{Conclusions}
In conclusion, we have demonstrated a Herriott cavity assisted vector magnetometer using dual orthogonal Bell-Bloom optical pumping beams, where we extract the field orientation from the phase delays between the modulated atomic signals and driving beams. We study the performance of this vector magnetometer in a bias field ranging from 100~mG to 500~mG, and demonstrate a characteristic field angle sensitivity better than 10~${\mu}$rad/Hz$^{1/2}$ above 10~Hz. Using the additional signal amplitude information, the magnetometer dead zone is limited to a region near the probe beam direction. We are working on eliminating this dead zone with a system consisting of two crossed Herriott cavities, and improving the system for better accuracy of the theoretical model and higher sensitivity on the magnetic field orientation.  In addition to the permanent electric dipole moment experiment~\cite{afach2015}, this vector magnetometer could also be applied in precision measurements such as the GNOME experiment~\cite{pospelov2013,pustelny2013,masiaroig2020}, where a high bandwidth vector magnetometer increases the detection sensitivity of the possible events coming through the magnetometer in the direction perpendicular to the bias field compared with scalar magnetometers~\cite{afach2018}.

\section{Acknowledgement}
We are grateful for the instrument support from Professor S.-M. Hu, Professor Z.-T. Lu, and Professor Y. Zhao. This work was supported by Natural Science Foundation of China (Grants No.11774329 and 11974329).


\begin{thebibliography}{37}%
\makeatletter
\providecommand \@ifxundefined [1]{%
 \@ifx{#1\undefined}
}%
\providecommand \@ifnum [1]{%
 \ifnum #1\expandafter \@firstoftwo
 \else \expandafter \@secondoftwo
 \fi
}%
\providecommand \@ifx [1]{%
 \ifx #1\expandafter \@firstoftwo
 \else \expandafter \@secondoftwo
 \fi
}%
\providecommand \natexlab [1]{#1}%
\providecommand \enquote  [1]{``#1''}%
\providecommand \bibnamefont  [1]{#1}%
\providecommand \bibfnamefont [1]{#1}%
\providecommand \citenamefont [1]{#1}%
\providecommand \href@noop [0]{\@secondoftwo}%
\providecommand \href [0]{\begingroup \@sanitize@url \@href}%
\providecommand \@href[1]{\@@startlink{#1}\@@href}%
\providecommand \@@href[1]{\endgroup#1\@@endlink}%
\providecommand \@sanitize@url [0]{\catcode `\\12\catcode `\$12\catcode
  `\&12\catcode `\#12\catcode `\^12\catcode `\_12\catcode `\%12\relax}%
\providecommand \@@startlink[1]{}%
\providecommand \@@endlink[0]{}%
\providecommand \url  [0]{\begingroup\@sanitize@url \@url }%
\providecommand \@url [1]{\endgroup\@href {#1}{\urlprefix }}%
\providecommand \urlprefix  [0]{URL }%
\providecommand \Eprint [0]{\href }%
\providecommand \doibase [0]{http://dx.doi.org/}%
\providecommand \selectlanguage [0]{\@gobble}%
\providecommand \bibinfo  [0]{\@secondoftwo}%
\providecommand \bibfield  [0]{\@secondoftwo}%
\providecommand \translation [1]{[#1]}%
\providecommand \BibitemOpen [0]{}%
\providecommand \bibitemStop [0]{}%
\providecommand \bibitemNoStop [0]{.\EOS\space}%
\providecommand \EOS [0]{\spacefactor3000\relax}%
\providecommand \BibitemShut  [1]{\csname bibitem#1\endcsname}%
\let\auto@bib@innerbib\@empty
\bibitem [{\citenamefont {Budker}\ and\ \citenamefont
  {Romalis}(2007)}]{budker2007}%
  \BibitemOpen
  \bibfield  {author} {\bibinfo {author} {\bibfnamefont {D.}~\bibnamefont
  {Budker}}\ and\ \bibinfo {author} {\bibfnamefont {M.}~\bibnamefont
  {Romalis}},\ }\href@noop {} {\bibfield  {journal} {\bibinfo  {journal} {Nat.
  Phys.}\ }\textbf {\bibinfo {volume} {3}},\ \bibinfo {pages} {227} (\bibinfo
  {year} {2007})}\BibitemShut {NoStop}%
\bibitem [{\citenamefont {Griffith}\ \emph {et~al.}(2009)\citenamefont
  {Griffith}, \citenamefont {Swallows}, \citenamefont {Loftus}, \citenamefont
  {Romalis}, \citenamefont {Heckel},\ and\ \citenamefont
  {Fortson}}]{griffith2009}%
  \BibitemOpen
  \bibfield  {author} {\bibinfo {author} {\bibfnamefont {W.~C.}\ \bibnamefont
  {Griffith}}, \bibinfo {author} {\bibfnamefont {M.~D.}\ \bibnamefont
  {Swallows}}, \bibinfo {author} {\bibfnamefont {T.~H.}\ \bibnamefont
  {Loftus}}, \bibinfo {author} {\bibfnamefont {M.~V.}\ \bibnamefont {Romalis}},
  \bibinfo {author} {\bibfnamefont {B.~R.}\ \bibnamefont {Heckel}}, \ and\
  \bibinfo {author} {\bibfnamefont {E.~N.}\ \bibnamefont {Fortson}},\ }\href
  {\doibase 10.1103/PhysRevLett.102.101601} {\bibfield  {journal} {\bibinfo
  {journal} {Phys. Rev. Lett.}\ }\textbf {\bibinfo {volume} {102}},\ \bibinfo
  {pages} {101601} (\bibinfo {year} {2009})}\BibitemShut {NoStop}%
\bibitem [{\citenamefont {Pospelov}\ \emph {et~al.}(2013)\citenamefont
  {Pospelov}, \citenamefont {Pustelny}, \citenamefont {Ledbetter},
  \citenamefont {Kimball}, \citenamefont {Gawlik},\ and\ \citenamefont
  {Budker}}]{pospelov2013}%
  \BibitemOpen
  \bibfield  {author} {\bibinfo {author} {\bibfnamefont {M.}~\bibnamefont
  {Pospelov}}, \bibinfo {author} {\bibfnamefont {S.}~\bibnamefont {Pustelny}},
  \bibinfo {author} {\bibfnamefont {M.~P.}\ \bibnamefont {Ledbetter}}, \bibinfo
  {author} {\bibfnamefont {D.~F.~J.}\ \bibnamefont {Kimball}}, \bibinfo
  {author} {\bibfnamefont {W.}~\bibnamefont {Gawlik}}, \ and\ \bibinfo {author}
  {\bibfnamefont {D.}~\bibnamefont {Budker}},\ }\href {\doibase
  10.1103/PhysRevLett.110.021803} {\bibfield  {journal} {\bibinfo  {journal}
  {Phys. Rev. Lett.}\ }\textbf {\bibinfo {volume} {110}},\ \bibinfo {pages}
  {021803} (\bibinfo {year} {2013})}\BibitemShut {NoStop}%
\bibitem [{\citenamefont {Knappe}\ \emph {et~al.}(2010)\citenamefont {Knappe},
  \citenamefont {Sander}, \citenamefont {Kosch}, \citenamefont {Wiekhorst},
  \citenamefont {Kitching},\ and\ \citenamefont {Trahms}}]{knappe2010}%
  \BibitemOpen
  \bibfield  {author} {\bibinfo {author} {\bibfnamefont {S.}~\bibnamefont
  {Knappe}}, \bibinfo {author} {\bibfnamefont {T.~H.}\ \bibnamefont {Sander}},
  \bibinfo {author} {\bibfnamefont {O.}~\bibnamefont {Kosch}}, \bibinfo
  {author} {\bibfnamefont {F.}~\bibnamefont {Wiekhorst}}, \bibinfo {author}
  {\bibfnamefont {J.}~\bibnamefont {Kitching}}, \ and\ \bibinfo {author}
  {\bibfnamefont {L.}~\bibnamefont {Trahms}},\ }\href {\doibase
  10.1063/1.3491548} {\bibfield  {journal} {\bibinfo  {journal} {Appl. Phys.
  Lett.}\ }\textbf {\bibinfo {volume} {97}},\ \bibinfo {pages} {133703}
  (\bibinfo {year} {2010})}\BibitemShut {NoStop}%
\bibitem [{\citenamefont {Maser}\ \emph {et~al.}(2011)\citenamefont {Maser},
  \citenamefont {Pandey}, \citenamefont {Ring}, \citenamefont {Ledbetter},
  \citenamefont {Knappe}, \citenamefont {Kitching},\ and\ \citenamefont
  {Budker}}]{maser2011}%
  \BibitemOpen
  \bibfield  {author} {\bibinfo {author} {\bibfnamefont {D.}~\bibnamefont
  {Maser}}, \bibinfo {author} {\bibfnamefont {S.}~\bibnamefont {Pandey}},
  \bibinfo {author} {\bibfnamefont {H.}~\bibnamefont {Ring}}, \bibinfo {author}
  {\bibfnamefont {M.~P.}\ \bibnamefont {Ledbetter}}, \bibinfo {author}
  {\bibfnamefont {S.}~\bibnamefont {Knappe}}, \bibinfo {author} {\bibfnamefont
  {J.}~\bibnamefont {Kitching}}, \ and\ \bibinfo {author} {\bibfnamefont
  {D.}~\bibnamefont {Budker}},\ }\href@noop {} {\bibfield  {journal} {\bibinfo
  {journal} {Rev. Sci. Instrum.}\ }\textbf {\bibinfo {volume} {82}},\ \bibinfo
  {pages} {086112} (\bibinfo {year} {2011})}\BibitemShut {NoStop}%
\bibitem [{\citenamefont {Hu}\ \emph {et~al.}(2020)\citenamefont {Hu},
  \citenamefont {Iwata}, \citenamefont {Mohammadi}, \citenamefont {Silletta},
  \citenamefont {Wickenbrock}, \citenamefont {Blanchard}, \citenamefont
  {Budker},\ and\ \citenamefont {Jerschow}}]{hu2019}%
  \BibitemOpen
  \bibfield  {author} {\bibinfo {author} {\bibfnamefont {Y.}~\bibnamefont
  {Hu}}, \bibinfo {author} {\bibfnamefont {G.~Z.}\ \bibnamefont {Iwata}},
  \bibinfo {author} {\bibfnamefont {M.}~\bibnamefont {Mohammadi}}, \bibinfo
  {author} {\bibfnamefont {E.~V.}\ \bibnamefont {Silletta}}, \bibinfo {author}
  {\bibfnamefont {A.}~\bibnamefont {Wickenbrock}}, \bibinfo {author}
  {\bibfnamefont {J.~W.}\ \bibnamefont {Blanchard}}, \bibinfo {author}
  {\bibfnamefont {D.}~\bibnamefont {Budker}}, \ and\ \bibinfo {author}
  {\bibfnamefont {A.}~\bibnamefont {Jerschow}},\ }\href@noop {} {\bibfield
  {journal} {\bibinfo  {journal} {PNAS}\ }\textbf {\bibinfo {volume} {117}},\
  \bibinfo {pages} {10667} (\bibinfo {year} {2020})}\BibitemShut {NoStop}%
\bibitem [{\citenamefont {Prouty}\ \emph {et~al.}(2013)\citenamefont {Prouty},
  \citenamefont {Johnson}, \citenamefont {Hrvoic},\ and\ \citenamefont
  {Vershovskiy}}]{prouty2013}%
  \BibitemOpen
  \bibfield  {author} {\bibinfo {author} {\bibfnamefont {M.~D.}\ \bibnamefont
  {Prouty}}, \bibinfo {author} {\bibfnamefont {R.}~\bibnamefont {Johnson}},
  \bibinfo {author} {\bibfnamefont {I.}~\bibnamefont {Hrvoic}}, \ and\ \bibinfo
  {author} {\bibfnamefont {A.~K.}\ \bibnamefont {Vershovskiy}},\ }in\
  \href@noop {} {\emph {\bibinfo {booktitle} {Optical Magnetometry}}}\
  (\bibinfo  {publisher} {Cambridge University Press},\ \bibinfo {year}
  {2013})\ pp.\ \bibinfo {pages} {369--386}\BibitemShut {NoStop}%
\bibitem [{\citenamefont {Kim}\ \emph {et~al.}(2014)\citenamefont {Kim},
  \citenamefont {Begus}, \citenamefont {Xia}, \citenamefont {Lee},
  \citenamefont {Jazbinsek}, \citenamefont {Trontelj},\ and\ \citenamefont
  {Romalis}}]{kim2014}%
  \BibitemOpen
  \bibfield  {author} {\bibinfo {author} {\bibfnamefont {K.}~\bibnamefont
  {Kim}}, \bibinfo {author} {\bibfnamefont {S.}~\bibnamefont {Begus}}, \bibinfo
  {author} {\bibfnamefont {H.}~\bibnamefont {Xia}}, \bibinfo {author}
  {\bibfnamefont {S.-K.}\ \bibnamefont {Lee}}, \bibinfo {author} {\bibfnamefont
  {V.}~\bibnamefont {Jazbinsek}}, \bibinfo {author} {\bibfnamefont
  {Z.}~\bibnamefont {Trontelj}}, \ and\ \bibinfo {author} {\bibfnamefont
  {M.~V.}\ \bibnamefont {Romalis}},\ }\href@noop {} {\bibfield  {journal}
  {\bibinfo  {journal} {NeuroImage}\ }\textbf {\bibinfo {volume} {89}},\
  \bibinfo {pages} {143} (\bibinfo {year} {2014})}\BibitemShut {NoStop}%
\bibitem [{\citenamefont {Alem}\ \emph {et~al.}(2015)\citenamefont {Alem},
  \citenamefont {Sander}, \citenamefont {Mhaskar}, \citenamefont {LeBlanc},
  \citenamefont {Eswaran}, \citenamefont {Steinhoff}, \citenamefont {Okada},
  \citenamefont {Kitching}, \citenamefont {Trahms},\ and\ \citenamefont
  {Knappe}}]{alem2015}%
  \BibitemOpen
  \bibfield  {author} {\bibinfo {author} {\bibfnamefont {O.}~\bibnamefont
  {Alem}}, \bibinfo {author} {\bibfnamefont {T.~H.}\ \bibnamefont {Sander}},
  \bibinfo {author} {\bibfnamefont {R.}~\bibnamefont {Mhaskar}}, \bibinfo
  {author} {\bibfnamefont {J.}~\bibnamefont {LeBlanc}}, \bibinfo {author}
  {\bibfnamefont {H.}~\bibnamefont {Eswaran}}, \bibinfo {author} {\bibfnamefont
  {U.}~\bibnamefont {Steinhoff}}, \bibinfo {author} {\bibfnamefont
  {Y.}~\bibnamefont {Okada}}, \bibinfo {author} {\bibfnamefont
  {J.}~\bibnamefont {Kitching}}, \bibinfo {author} {\bibfnamefont
  {L.}~\bibnamefont {Trahms}}, \ and\ \bibinfo {author} {\bibfnamefont
  {S.}~\bibnamefont {Knappe}},\ }\href@noop {} {\bibfield  {journal} {\bibinfo
  {journal} {Phys. Med. Biol.}\ }\textbf {\bibinfo {volume} {60}},\ \bibinfo
  {pages} {4797} (\bibinfo {year} {2015})}\BibitemShut {NoStop}%
\bibitem [{\citenamefont {Boto}\ \emph {et~al.}(2018)\citenamefont {Boto},
  \citenamefont {Holmes}, \citenamefont {Leggett}, \citenamefont {Roberts},
  \citenamefont {Shah}, \citenamefont {Meyer}, \citenamefont {Mu{\~n}oz},
  \citenamefont {Mullinger}, \citenamefont {Tierney}, \citenamefont {Bestmann}
  \emph {et~al.}}]{boto18}%
  \BibitemOpen
  \bibfield  {author} {\bibinfo {author} {\bibfnamefont {E.}~\bibnamefont
  {Boto}}, \bibinfo {author} {\bibfnamefont {N.}~\bibnamefont {Holmes}},
  \bibinfo {author} {\bibfnamefont {J.}~\bibnamefont {Leggett}}, \bibinfo
  {author} {\bibfnamefont {G.}~\bibnamefont {Roberts}}, \bibinfo {author}
  {\bibfnamefont {V.}~\bibnamefont {Shah}}, \bibinfo {author} {\bibfnamefont
  {S.~S.}\ \bibnamefont {Meyer}}, \bibinfo {author} {\bibfnamefont {L.~D.}\
  \bibnamefont {Mu{\~n}oz}}, \bibinfo {author} {\bibfnamefont {K.~J.}\
  \bibnamefont {Mullinger}}, \bibinfo {author} {\bibfnamefont {T.~M.}\
  \bibnamefont {Tierney}}, \bibinfo {author} {\bibfnamefont {S.}~\bibnamefont
  {Bestmann}},  \emph {et~al.},\ }\href@noop {} {\bibfield  {journal} {\bibinfo
   {journal} {Nature}\ }\textbf {\bibinfo {volume} {555}},\ \bibinfo {pages}
  {657} (\bibinfo {year} {2018})}\BibitemShut {NoStop}%
\bibitem [{\citenamefont {Lee}\ \emph {et~al.}(2006)\citenamefont {Lee},
  \citenamefont {Sauer}, \citenamefont {Seltzer}, \citenamefont {Alem},\ and\
  \citenamefont {Romalis}}]{Lee2006}%
  \BibitemOpen
  \bibfield  {author} {\bibinfo {author} {\bibfnamefont {S.-K.}\ \bibnamefont
  {Lee}}, \bibinfo {author} {\bibfnamefont {K.~L.}\ \bibnamefont {Sauer}},
  \bibinfo {author} {\bibfnamefont {S.~J.}\ \bibnamefont {Seltzer}}, \bibinfo
  {author} {\bibfnamefont {O.}~\bibnamefont {Alem}}, \ and\ \bibinfo {author}
  {\bibfnamefont {M.~V.}\ \bibnamefont {Romalis}},\ }\href {\doibase
  10.1063/1.2390643} {\bibfield  {journal} {\bibinfo  {journal} {Appl. Phys.
  Lett.}\ }\textbf {\bibinfo {volume} {89}},\ \bibinfo {pages} {214106}
  (\bibinfo {year} {2006})}\BibitemShut {NoStop}%
\bibitem [{\citenamefont {Dang}\ \emph {et~al.}(2010)\citenamefont {Dang},
  \citenamefont {Maloof},\ and\ \citenamefont {Romalis}}]{dang2010}%
  \BibitemOpen
  \bibfield  {author} {\bibinfo {author} {\bibfnamefont {H.~B.}\ \bibnamefont
  {Dang}}, \bibinfo {author} {\bibfnamefont {A.~C.}\ \bibnamefont {Maloof}}, \
  and\ \bibinfo {author} {\bibfnamefont {M.~V.}\ \bibnamefont {Romalis}},\
  }\href {\doibase 10.1063/1.3491215} {\bibfield  {journal} {\bibinfo
  {journal} {Appl. Phys. Lett.}\ }\textbf {\bibinfo {volume} {97}},\ \bibinfo
  {pages} {151110} (\bibinfo {year} {2010})}\BibitemShut {NoStop}%
\bibitem [{\citenamefont {Liew}\ \emph {et~al.}(2004)\citenamefont {Liew},
  \citenamefont {Knappe}, \citenamefont {Moreland}, \citenamefont {Robinson},
  \citenamefont {Hollberg},\ and\ \citenamefont {Kitching}}]{liew2004}%
  \BibitemOpen
  \bibfield  {author} {\bibinfo {author} {\bibfnamefont {L.-A.}\ \bibnamefont
  {Liew}}, \bibinfo {author} {\bibfnamefont {S.}~\bibnamefont {Knappe}},
  \bibinfo {author} {\bibfnamefont {J.}~\bibnamefont {Moreland}}, \bibinfo
  {author} {\bibfnamefont {H.}~\bibnamefont {Robinson}}, \bibinfo {author}
  {\bibfnamefont {L.}~\bibnamefont {Hollberg}}, \ and\ \bibinfo {author}
  {\bibfnamefont {J.}~\bibnamefont {Kitching}},\ }\href {\doibase
  10.1063/1.1691490} {\bibfield  {journal} {\bibinfo  {journal} {Appl. Phys.
  Lett.}\ }\textbf {\bibinfo {volume} {84}},\ \bibinfo {pages} {2694} (\bibinfo
  {year} {2004})}\BibitemShut {NoStop}%
\bibitem [{\citenamefont {Mhaskar}\ \emph {et~al.}(2012)\citenamefont
  {Mhaskar}, \citenamefont {Knappe},\ and\ \citenamefont
  {Kitching}}]{mhaskar12}%
  \BibitemOpen
  \bibfield  {author} {\bibinfo {author} {\bibfnamefont {R.}~\bibnamefont
  {Mhaskar}}, \bibinfo {author} {\bibfnamefont {S.}~\bibnamefont {Knappe}}, \
  and\ \bibinfo {author} {\bibfnamefont {J.}~\bibnamefont {Kitching}},\
  }\href@noop {} {\bibfield  {journal} {\bibinfo  {journal} {Appl. Phys.
  Lett.}\ }\textbf {\bibinfo {volume} {101}},\ \bibinfo {pages} {241105}
  (\bibinfo {year} {2012})}\BibitemShut {NoStop}%
\bibitem [{\citenamefont {Kitching}(2018)}]{kitching2018}%
  \BibitemOpen
  \bibfield  {author} {\bibinfo {author} {\bibfnamefont {J.}~\bibnamefont
  {Kitching}},\ }\href {\doibase 10.1063/1.5026238} {\bibfield  {journal}
  {\bibinfo  {journal} {Appl. Phys. Rev.}\ }\textbf {\bibinfo {volume} {5}},\
  \bibinfo {pages} {031302} (\bibinfo {year} {2018})}\BibitemShut {NoStop}%
\bibitem [{\citenamefont {Li}\ \emph {et~al.}(2011)\citenamefont {Li},
  \citenamefont {Vachaspati}, \citenamefont {Sheng}, \citenamefont {Dural},\
  and\ \citenamefont {Romalis}}]{li2011}%
  \BibitemOpen
  \bibfield  {author} {\bibinfo {author} {\bibfnamefont {S.}~\bibnamefont
  {Li}}, \bibinfo {author} {\bibfnamefont {P.}~\bibnamefont {Vachaspati}},
  \bibinfo {author} {\bibfnamefont {D.}~\bibnamefont {Sheng}}, \bibinfo
  {author} {\bibfnamefont {N.}~\bibnamefont {Dural}}, \ and\ \bibinfo {author}
  {\bibfnamefont {M.~V.}\ \bibnamefont {Romalis}},\ }\href {\doibase
  10.1103/PhysRevA.84.061403} {\bibfield  {journal} {\bibinfo  {journal} {Phys.
  Rev. A}\ }\textbf {\bibinfo {volume} {84}},\ \bibinfo {pages} {061403}
  (\bibinfo {year} {2011})}\BibitemShut {NoStop}%
\bibitem [{\citenamefont {Jensen}\ \emph {et~al.}(2014)\citenamefont {Jensen},
  \citenamefont {Leefer}, \citenamefont {Jarmola}, \citenamefont {Dumeige},
  \citenamefont {Acosta}, \citenamefont {Kehayias}, \citenamefont {Patton},\
  and\ \citenamefont {Budker}}]{jensen2014}%
  \BibitemOpen
  \bibfield  {author} {\bibinfo {author} {\bibfnamefont {K.}~\bibnamefont
  {Jensen}}, \bibinfo {author} {\bibfnamefont {N.}~\bibnamefont {Leefer}},
  \bibinfo {author} {\bibfnamefont {A.}~\bibnamefont {Jarmola}}, \bibinfo
  {author} {\bibfnamefont {Y.}~\bibnamefont {Dumeige}}, \bibinfo {author}
  {\bibfnamefont {V.~M.}\ \bibnamefont {Acosta}}, \bibinfo {author}
  {\bibfnamefont {P.}~\bibnamefont {Kehayias}}, \bibinfo {author}
  {\bibfnamefont {B.}~\bibnamefont {Patton}}, \ and\ \bibinfo {author}
  {\bibfnamefont {D.}~\bibnamefont {Budker}},\ }\href {\doibase
  10.1103/PhysRevLett.112.160802} {\bibfield  {journal} {\bibinfo  {journal}
  {Phys. Rev. Lett.}\ }\textbf {\bibinfo {volume} {112}},\ \bibinfo {pages}
  {160802} (\bibinfo {year} {2014})}\BibitemShut {NoStop}%
\bibitem [{\citenamefont {Clevenson}\ \emph {et~al.}(2015)\citenamefont
  {Clevenson}, \citenamefont {Trusheim}, \citenamefont {Teale}, \citenamefont
  {Schröder}, \citenamefont {Braje},\ and\ \citenamefont
  {Englund}}]{clevenson2015}%
  \BibitemOpen
  \bibfield  {author} {\bibinfo {author} {\bibfnamefont {H.}~\bibnamefont
  {Clevenson}}, \bibinfo {author} {\bibfnamefont {M.~E.}\ \bibnamefont
  {Trusheim}}, \bibinfo {author} {\bibfnamefont {C.}~\bibnamefont {Teale}},
  \bibinfo {author} {\bibfnamefont {T.}~\bibnamefont {Schröder}}, \bibinfo
  {author} {\bibfnamefont {D.}~\bibnamefont {Braje}}, \ and\ \bibinfo {author}
  {\bibfnamefont {D.}~\bibnamefont {Englund}},\ }\href {\doibase
  10.1038/nphys3291} {\bibfield  {journal} {\bibinfo  {journal} {Nat. Phys.}\
  }\textbf {\bibinfo {volume} {11}},\ \bibinfo {pages} {393} (\bibinfo {year}
  {2015})}\BibitemShut {NoStop}%
\bibitem [{\citenamefont {Shi}\ \emph {et~al.}(2013)\citenamefont {Shi},
  \citenamefont {Ikäläinen}, \citenamefont {Vaara},\ and\ \citenamefont
  {Romalis}}]{shi2013}%
  \BibitemOpen
  \bibfield  {author} {\bibinfo {author} {\bibfnamefont {J.}~\bibnamefont
  {Shi}}, \bibinfo {author} {\bibfnamefont {S.}~\bibnamefont {Ikäläinen}},
  \bibinfo {author} {\bibfnamefont {J.}~\bibnamefont {Vaara}}, \ and\ \bibinfo
  {author} {\bibfnamefont {M.~V.}\ \bibnamefont {Romalis}},\ }\href {\doibase
  10.1021/jz3018539} {\bibfield  {journal} {\bibinfo  {journal} {J. Phys. Chem.
  Lett.}\ }\textbf {\bibinfo {volume} {4}},\ \bibinfo {pages} {437} (\bibinfo
  {year} {2013})}\BibitemShut {NoStop}%
\bibitem [{\citenamefont {Cooper}\ \emph {et~al.}(2016)\citenamefont {Cooper},
  \citenamefont {Prescott}, \citenamefont {Matz}, \citenamefont {Sauer},
  \citenamefont {Dural}, \citenamefont {Romalis}, \citenamefont {Foley},
  \citenamefont {Kornack}, \citenamefont {Monti},\ and\ \citenamefont
  {Okamitsu}}]{cooper2016}%
  \BibitemOpen
  \bibfield  {author} {\bibinfo {author} {\bibfnamefont {R.~J.}\ \bibnamefont
  {Cooper}}, \bibinfo {author} {\bibfnamefont {D.~W.}\ \bibnamefont
  {Prescott}}, \bibinfo {author} {\bibfnamefont {P.}~\bibnamefont {Matz}},
  \bibinfo {author} {\bibfnamefont {K.~L.}\ \bibnamefont {Sauer}}, \bibinfo
  {author} {\bibfnamefont {N.}~\bibnamefont {Dural}}, \bibinfo {author}
  {\bibfnamefont {M.~V.}\ \bibnamefont {Romalis}}, \bibinfo {author}
  {\bibfnamefont {E.~L.}\ \bibnamefont {Foley}}, \bibinfo {author}
  {\bibfnamefont {T.~W.}\ \bibnamefont {Kornack}}, \bibinfo {author}
  {\bibfnamefont {M.}~\bibnamefont {Monti}}, \ and\ \bibinfo {author}
  {\bibfnamefont {J.}~\bibnamefont {Okamitsu}},\ }\href {\doibase
  10.1103/PhysRevApplied.6.064014} {\bibfield  {journal} {\bibinfo  {journal}
  {Phys. Rev. Applied}\ }\textbf {\bibinfo {volume} {6}},\ \bibinfo {pages}
  {064014} (\bibinfo {year} {2016})}\BibitemShut {NoStop}%
\bibitem [{\citenamefont {Sheng}\ \emph {et~al.}(2013)\citenamefont {Sheng},
  \citenamefont {Li}, \citenamefont {Dural},\ and\ \citenamefont
  {Romalis}}]{sheng2013}%
  \BibitemOpen
  \bibfield  {author} {\bibinfo {author} {\bibfnamefont {D.}~\bibnamefont
  {Sheng}}, \bibinfo {author} {\bibfnamefont {S.}~\bibnamefont {Li}}, \bibinfo
  {author} {\bibfnamefont {N.}~\bibnamefont {Dural}}, \ and\ \bibinfo {author}
  {\bibfnamefont {M.~V.}\ \bibnamefont {Romalis}},\ }\href {\doibase
  10.1103/PhysRevLett.110.160802} {\bibfield  {journal} {\bibinfo  {journal}
  {Phys. Rev. Lett.}\ }\textbf {\bibinfo {volume} {110}},\ \bibinfo {pages}
  {160802} (\bibinfo {year} {2013})}\BibitemShut {NoStop}%
\bibitem [{\citenamefont {Limes}\ \emph {et~al.}(2020)\citenamefont {Limes},
  \citenamefont {Foley}, \citenamefont {Kornack}, \citenamefont {Caliga},
  \citenamefont {McBride}, \citenamefont {Braun}, \citenamefont {Lee},
  \citenamefont {Lucivero},\ and\ \citenamefont {Romalis}}]{limes2020}%
  \BibitemOpen
  \bibfield  {author} {\bibinfo {author} {\bibfnamefont {M.~E.}\ \bibnamefont
  {Limes}}, \bibinfo {author} {\bibfnamefont {E.~L.}\ \bibnamefont {Foley}},
  \bibinfo {author} {\bibfnamefont {T.~W.}\ \bibnamefont {Kornack}}, \bibinfo
  {author} {\bibfnamefont {S.}~\bibnamefont {Caliga}}, \bibinfo {author}
  {\bibfnamefont {S.}~\bibnamefont {McBride}}, \bibinfo {author} {\bibfnamefont
  {A.}~\bibnamefont {Braun}}, \bibinfo {author} {\bibfnamefont
  {W.}~\bibnamefont {Lee}}, \bibinfo {author} {\bibfnamefont {V.~G.}\
  \bibnamefont {Lucivero}}, \ and\ \bibinfo {author} {\bibfnamefont {M.~V.}\
  \bibnamefont {Romalis}},\ }\href@noop {} {\bibfield  {journal} {\bibinfo
  {journal} {arXiv:2001.03534}\ } (\bibinfo {year} {2020})}\BibitemShut
  {NoStop}%
\bibitem [{\citenamefont {Afach}\ \emph {et~al.}(2015)\citenamefont {Afach},
  \citenamefont {Ban}, \citenamefont {Bison}, \citenamefont {Bodek},
  \citenamefont {Chowdhuri}, \citenamefont {Gruji\'{c}}, \citenamefont {Hayen},
  \citenamefont {H\'{e}laine}, \citenamefont {Kasprzak}, \citenamefont {Kirch},
  \citenamefont {Knowles}, \citenamefont {Koch} \emph {et~al.}}]{afach2015}%
  \BibitemOpen
  \bibfield  {author} {\bibinfo {author} {\bibfnamefont {S.}~\bibnamefont
  {Afach}}, \bibinfo {author} {\bibfnamefont {G.}~\bibnamefont {Ban}}, \bibinfo
  {author} {\bibfnamefont {G.}~\bibnamefont {Bison}}, \bibinfo {author}
  {\bibfnamefont {K.}~\bibnamefont {Bodek}}, \bibinfo {author} {\bibfnamefont
  {Z.}~\bibnamefont {Chowdhuri}}, \bibinfo {author} {\bibfnamefont {Z.~D.}\
  \bibnamefont {Gruji\'{c}}}, \bibinfo {author} {\bibfnamefont
  {L.}~\bibnamefont {Hayen}}, \bibinfo {author} {\bibfnamefont
  {V.}~\bibnamefont {H\'{e}laine}}, \bibinfo {author} {\bibfnamefont
  {M.}~\bibnamefont {Kasprzak}}, \bibinfo {author} {\bibfnamefont
  {K.}~\bibnamefont {Kirch}}, \bibinfo {author} {\bibfnamefont
  {P.}~\bibnamefont {Knowles}}, \bibinfo {author} {\bibfnamefont {H.-C.}\
  \bibnamefont {Koch}},  \emph {et~al.},\ }\href {\doibase
  10.1364/OE.23.022108} {\bibfield  {journal} {\bibinfo  {journal} {Opt. Exp.}\
  }\textbf {\bibinfo {volume} {23}},\ \bibinfo {pages} {22108} (\bibinfo {year}
  {2015})}\BibitemShut {NoStop}%
\bibitem [{\citenamefont {Patton}\ \emph {et~al.}(2014)\citenamefont {Patton},
  \citenamefont {Zhivun}, \citenamefont {Hovde},\ and\ \citenamefont
  {Budker}}]{patton2014}%
  \BibitemOpen
  \bibfield  {author} {\bibinfo {author} {\bibfnamefont {B.}~\bibnamefont
  {Patton}}, \bibinfo {author} {\bibfnamefont {E.}~\bibnamefont {Zhivun}},
  \bibinfo {author} {\bibfnamefont {D.~C.}\ \bibnamefont {Hovde}}, \ and\
  \bibinfo {author} {\bibfnamefont {D.}~\bibnamefont {Budker}},\ }\href
  {\doibase 10.1103/PhysRevLett.113.013001} {\bibfield  {journal} {\bibinfo
  {journal} {Phys. Rev. Lett.}\ }\textbf {\bibinfo {volume} {113}},\ \bibinfo
  {pages} {013001} (\bibinfo {year} {2014})}\BibitemShut {NoStop}%
\bibitem [{\citenamefont {Huang}\ \emph {et~al.}(2015)\citenamefont {Huang},
  \citenamefont {Dong}, \citenamefont {Hu}, \citenamefont {Chen},\ and\
  \citenamefont {Gao}}]{huang2015}%
  \BibitemOpen
  \bibfield  {author} {\bibinfo {author} {\bibfnamefont {H.}~\bibnamefont
  {Huang}}, \bibinfo {author} {\bibfnamefont {H.}~\bibnamefont {Dong}},
  \bibinfo {author} {\bibfnamefont {X.~Y.}\ \bibnamefont {Hu}}, \bibinfo
  {author} {\bibfnamefont {L.}~\bibnamefont {Chen}}, \ and\ \bibinfo {author}
  {\bibfnamefont {Y.}~\bibnamefont {Gao}},\ }\href@noop {} {\bibfield
  {journal} {\bibinfo  {journal} {Appl. Phys. Lett.}\ }\textbf {\bibinfo
  {volume} {107}},\ \bibinfo {pages} {182403} (\bibinfo {year}
  {2015})}\BibitemShut {NoStop}%
\bibitem [{\citenamefont {Huang}\ \emph {et~al.}(2016)\citenamefont {Huang},
  \citenamefont {Dong}, \citenamefont {Chen},\ and\ \citenamefont
  {Gao}}]{huang2016}%
  \BibitemOpen
  \bibfield  {author} {\bibinfo {author} {\bibfnamefont {H.}~\bibnamefont
  {Huang}}, \bibinfo {author} {\bibfnamefont {H.}~\bibnamefont {Dong}},
  \bibinfo {author} {\bibfnamefont {L.}~\bibnamefont {Chen}}, \ and\ \bibinfo
  {author} {\bibfnamefont {Y.}~\bibnamefont {Gao}},\ }\href@noop {} {\bibfield
  {journal} {\bibinfo  {journal} {Appl. Phys. Lett.}\ }\textbf {\bibinfo
  {volume} {109}},\ \bibinfo {pages} {062404} (\bibinfo {year}
  {2016})}\BibitemShut {NoStop}%
\bibitem [{\citenamefont {Ingleby}\ \emph {et~al.}(2018)\citenamefont
  {Ingleby}, \citenamefont {O'Dwyer}, \citenamefont {Griffin}, \citenamefont
  {Arnold},\ and\ \citenamefont {Riis}}]{ingleby2018}%
  \BibitemOpen
  \bibfield  {author} {\bibinfo {author} {\bibfnamefont {S.~J.}\ \bibnamefont
  {Ingleby}}, \bibinfo {author} {\bibfnamefont {C.}~\bibnamefont {O'Dwyer}},
  \bibinfo {author} {\bibfnamefont {P.~F.}\ \bibnamefont {Griffin}}, \bibinfo
  {author} {\bibfnamefont {A.~S.}\ \bibnamefont {Arnold}}, \ and\ \bibinfo
  {author} {\bibfnamefont {E.}~\bibnamefont {Riis}},\ }\href {\doibase
  10.1103/PhysRevApplied.10.034035} {\bibfield  {journal} {\bibinfo  {journal}
  {Phys. Rev. Applied}\ }\textbf {\bibinfo {volume} {10}},\ \bibinfo {pages}
  {034035} (\bibinfo {year} {2018})}\BibitemShut {NoStop}%
\bibitem [{\citenamefont {Pyragius}\ \emph {et~al.}(2019)\citenamefont
  {Pyragius}, \citenamefont {Florez},\ and\ \citenamefont
  {Fernholz}}]{pyragius2019}%
  \BibitemOpen
  \bibfield  {author} {\bibinfo {author} {\bibfnamefont {T.}~\bibnamefont
  {Pyragius}}, \bibinfo {author} {\bibfnamefont {H.~M.}\ \bibnamefont
  {Florez}}, \ and\ \bibinfo {author} {\bibfnamefont {T.}~\bibnamefont
  {Fernholz}},\ }\href {\doibase 10.1103/PhysRevA.100.023416} {\bibfield
  {journal} {\bibinfo  {journal} {Phys. Rev. A}\ }\textbf {\bibinfo {volume}
  {100}},\ \bibinfo {pages} {023416} (\bibinfo {year} {2019})}\BibitemShut
  {NoStop}%
\bibitem [{\citenamefont {Sun}\ \emph {et~al.}(2017)\citenamefont {Sun},
  \citenamefont {Huang}, \citenamefont {Huang}, \citenamefont {Wang},\ and\
  \citenamefont {Zhang}}]{sun2017}%
  \BibitemOpen
  \bibfield  {author} {\bibinfo {author} {\bibfnamefont {W.}~\bibnamefont
  {Sun}}, \bibinfo {author} {\bibfnamefont {Q.}~\bibnamefont {Huang}}, \bibinfo
  {author} {\bibfnamefont {Z.}~\bibnamefont {Huang}}, \bibinfo {author}
  {\bibfnamefont {P.}~\bibnamefont {Wang}}, \ and\ \bibinfo {author}
  {\bibfnamefont {J.}~\bibnamefont {Zhang}},\ }\href@noop {} {\bibfield
  {journal} {\bibinfo  {journal} {Chin. Phys. Lett.}\ }\textbf {\bibinfo
  {volume} {34}},\ \bibinfo {pages} {58501} (\bibinfo {year}
  {2017})}\BibitemShut {NoStop}%
\bibitem [{\citenamefont {Bell}\ and\ \citenamefont {Bloom}(1961)}]{bell1961}%
  \BibitemOpen
  \bibfield  {author} {\bibinfo {author} {\bibfnamefont {W.~E.}\ \bibnamefont
  {Bell}}\ and\ \bibinfo {author} {\bibfnamefont {A.~L.}\ \bibnamefont
  {Bloom}},\ }\href {\doibase 10.1103/PhysRevLett.6.280} {\bibfield  {journal}
  {\bibinfo  {journal} {Phys. Rev. Lett.}\ }\textbf {\bibinfo {volume} {6}},\
  \bibinfo {pages} {280} (\bibinfo {year} {1961})}\BibitemShut {NoStop}%
\bibitem [{\citenamefont {Silver}(2005)}]{silver2005}%
  \BibitemOpen
  \bibfield  {author} {\bibinfo {author} {\bibfnamefont {J.~A.}\ \bibnamefont
  {Silver}},\ }\href@noop {} {\bibfield  {journal} {\bibinfo  {journal} {App.
  Opt.}\ }\textbf {\bibinfo {volume} {44}},\ \bibinfo {pages} {6545} (\bibinfo
  {year} {2005})}\BibitemShut {NoStop}%
\bibitem [{\citenamefont {Hao}\ \emph {et~al.}(2019)\citenamefont {Hao},
  \citenamefont {Qiu}, \citenamefont {Sun}, \citenamefont {Zhu},\ and\
  \citenamefont {Sheng}}]{hao2019}%
  \BibitemOpen
  \bibfield  {author} {\bibinfo {author} {\bibfnamefont {C.-P.}\ \bibnamefont
  {Hao}}, \bibinfo {author} {\bibfnamefont {Z.-R.}\ \bibnamefont {Qiu}},
  \bibinfo {author} {\bibfnamefont {Q.}~\bibnamefont {Sun}}, \bibinfo {author}
  {\bibfnamefont {Y.}~\bibnamefont {Zhu}}, \ and\ \bibinfo {author}
  {\bibfnamefont {D.}~\bibnamefont {Sheng}},\ }\href {\doibase
  10.1103/PhysRevA.99.053417} {\bibfield  {journal} {\bibinfo  {journal} {Phys.
  Rev. A}\ }\textbf {\bibinfo {volume} {99}},\ \bibinfo {pages} {053417}
  (\bibinfo {year} {2019})}\BibitemShut {NoStop}%
\bibitem [{\citenamefont {Smullin}\ \emph {et~al.}(2009)\citenamefont
  {Smullin}, \citenamefont {Savukov}, \citenamefont {Vasilakis}, \citenamefont
  {Ghosh},\ and\ \citenamefont {Romalis}}]{smullin2009}%
  \BibitemOpen
  \bibfield  {author} {\bibinfo {author} {\bibfnamefont {S.~J.}\ \bibnamefont
  {Smullin}}, \bibinfo {author} {\bibfnamefont {I.~M.}\ \bibnamefont
  {Savukov}}, \bibinfo {author} {\bibfnamefont {G.}~\bibnamefont {Vasilakis}},
  \bibinfo {author} {\bibfnamefont {R.~K.}\ \bibnamefont {Ghosh}}, \ and\
  \bibinfo {author} {\bibfnamefont {M.~V.}\ \bibnamefont {Romalis}},\ }\href
  {\doibase 10.1103/PhysRevA.80.033420} {\bibfield  {journal} {\bibinfo
  {journal} {Phys. Rev. A}\ }\textbf {\bibinfo {volume} {80}},\ \bibinfo
  {pages} {033420} (\bibinfo {year} {2009})}\BibitemShut {NoStop}%
\bibitem [{\citenamefont {Appelt}\ \emph {et~al.}(1998)\citenamefont {Appelt},
  \citenamefont {Baranga}, \citenamefont {Erickson}, \citenamefont {Romalis},
  \citenamefont {Young},\ and\ \citenamefont {Happer}}]{appelt98}%
  \BibitemOpen
  \bibfield  {author} {\bibinfo {author} {\bibfnamefont {S.}~\bibnamefont
  {Appelt}}, \bibinfo {author} {\bibfnamefont {A.~B.}\ \bibnamefont {Baranga}},
  \bibinfo {author} {\bibfnamefont {C.~J.}\ \bibnamefont {Erickson}}, \bibinfo
  {author} {\bibfnamefont {M.}~\bibnamefont {Romalis}}, \bibinfo {author}
  {\bibfnamefont {A.~R.}\ \bibnamefont {Young}}, \ and\ \bibinfo {author}
  {\bibfnamefont {W.}~\bibnamefont {Happer}},\ }\href@noop {} {\bibfield
  {journal} {\bibinfo  {journal} {Phys. Rev. A}\ }\textbf {\bibinfo {volume}
  {58}},\ \bibinfo {pages} {1412} (\bibinfo {year} {1998})}\BibitemShut
  {NoStop}%
\bibitem [{\citenamefont {Pustelny}\ \emph {et~al.}(2013)\citenamefont
  {Pustelny}, \citenamefont {Kimball}, \citenamefont {Pankow}, \citenamefont
  {Ledbetter}, \citenamefont {Wlodarczyk}, \citenamefont {Wcislo},
  \citenamefont {Pospelov}, \citenamefont {Smith}, \citenamefont {Read},
  \citenamefont {Gawlik} \emph {et~al.}}]{pustelny2013}%
  \BibitemOpen
  \bibfield  {author} {\bibinfo {author} {\bibfnamefont {S.}~\bibnamefont
  {Pustelny}}, \bibinfo {author} {\bibfnamefont {D.~F.}\ \bibnamefont
  {Kimball}}, \bibinfo {author} {\bibfnamefont {C.}~\bibnamefont {Pankow}},
  \bibinfo {author} {\bibfnamefont {M.~P.}\ \bibnamefont {Ledbetter}}, \bibinfo
  {author} {\bibfnamefont {P.}~\bibnamefont {Wlodarczyk}}, \bibinfo {author}
  {\bibfnamefont {P.}~\bibnamefont {Wcislo}}, \bibinfo {author} {\bibfnamefont
  {M.}~\bibnamefont {Pospelov}}, \bibinfo {author} {\bibfnamefont {J.~R.}\
  \bibnamefont {Smith}}, \bibinfo {author} {\bibfnamefont {J.}~\bibnamefont
  {Read}}, \bibinfo {author} {\bibfnamefont {W.}~\bibnamefont {Gawlik}},  \emph
  {et~al.},\ }\href@noop {} {\bibfield  {journal} {\bibinfo  {journal} {Ann.
  Phys.}\ }\textbf {\bibinfo {volume} {525}},\ \bibinfo {pages} {659} (\bibinfo
  {year} {2013})}\BibitemShut {NoStop}%
\bibitem [{\citenamefont {Masia-Roig}\ \emph {et~al.}(2020)\citenamefont
  {Masia-Roig}, \citenamefont {Smiga}, \citenamefont {Budker}, \citenamefont
  {Dumont}, \citenamefont {Grujic}, \citenamefont {Kim}, \citenamefont
  {Kimball}, \citenamefont {Lebedev}, \citenamefont {Monroy}, \citenamefont
  {Pustelny}, \citenamefont {Scholtes}, \citenamefont {Segura}, \citenamefont
  {Semertzidis}, \citenamefont {Shin}, \citenamefont {Stalnaker}, \citenamefont
  {Sulai}, \citenamefont {Weis},\ and\ \citenamefont
  {Wickenbrock}}]{masiaroig2020}%
  \BibitemOpen
  \bibfield  {author} {\bibinfo {author} {\bibfnamefont {H.}~\bibnamefont
  {Masia-Roig}}, \bibinfo {author} {\bibfnamefont {J.~A.}\ \bibnamefont
  {Smiga}}, \bibinfo {author} {\bibfnamefont {D.}~\bibnamefont {Budker}},
  \bibinfo {author} {\bibfnamefont {V.}~\bibnamefont {Dumont}}, \bibinfo
  {author} {\bibfnamefont {Z.}~\bibnamefont {Grujic}}, \bibinfo {author}
  {\bibfnamefont {D.}~\bibnamefont {Kim}}, \bibinfo {author} {\bibfnamefont
  {D.~F.~J.}\ \bibnamefont {Kimball}}, \bibinfo {author} {\bibfnamefont
  {V.}~\bibnamefont {Lebedev}}, \bibinfo {author} {\bibfnamefont
  {M.}~\bibnamefont {Monroy}}, \bibinfo {author} {\bibfnamefont
  {S.}~\bibnamefont {Pustelny}}, \bibinfo {author} {\bibfnamefont
  {T.}~\bibnamefont {Scholtes}}, \bibinfo {author} {\bibfnamefont {P.~C.}\
  \bibnamefont {Segura}}, \bibinfo {author} {\bibfnamefont {Y.~K.}\
  \bibnamefont {Semertzidis}}, \bibinfo {author} {\bibfnamefont {Y.~C.}\
  \bibnamefont {Shin}}, \bibinfo {author} {\bibfnamefont {J.~E.}\ \bibnamefont
  {Stalnaker}}, \bibinfo {author} {\bibfnamefont {I.}~\bibnamefont {Sulai}},
  \bibinfo {author} {\bibfnamefont {A.}~\bibnamefont {Weis}}, \ and\ \bibinfo
  {author} {\bibfnamefont {A.}~\bibnamefont {Wickenbrock}},\ }\href {\doibase
  https://doi.org/10.1016/j.dark.2020.100494} {\bibfield  {journal} {\bibinfo
  {journal} {Physics of the Dark Universe}\ }\textbf {\bibinfo {volume} {28}},\
  \bibinfo {pages} {100494} (\bibinfo {year} {2020})}\BibitemShut {NoStop}%
\bibitem [{\citenamefont {Afach}\ \emph {et~al.}(2018)\citenamefont {Afach},
  \citenamefont {Budker}, \citenamefont {DeCamp}, \citenamefont {Dumont},
  \citenamefont {Grujić}, \citenamefont {Guo}, \citenamefont
  {Jackson Kimball}, \citenamefont {Kornack}, \citenamefont {Lebedev},
  \citenamefont {Li}, \citenamefont {Masia-Roig}, \citenamefont {Nix},
  \citenamefont {Padniuk}, \citenamefont {Palm}, \citenamefont {Pankow},
  \citenamefont {Penaflor}, \citenamefont {Peng}, \citenamefont {Pustelny},
  \citenamefont {Scholtes}, \citenamefont {Smiga}, \citenamefont {Stalnaker},
  \citenamefont {Weis}, \citenamefont {Wickenbrock},\ and\ \citenamefont
  {Wurm}}]{afach2018}%
  \BibitemOpen
  \bibfield  {author} {\bibinfo {author} {\bibfnamefont {S.}~\bibnamefont
  {Afach}}, \bibinfo {author} {\bibfnamefont {D.}~\bibnamefont {Budker}},
  \bibinfo {author} {\bibfnamefont {G.}~\bibnamefont {DeCamp}}, \bibinfo
  {author} {\bibfnamefont {V.}~\bibnamefont {Dumont}}, \bibinfo {author}
  {\bibfnamefont {Z.}~\bibnamefont {Grujić}}, \bibinfo {author} {\bibfnamefont
  {H.}~\bibnamefont {Guo}}, \bibinfo {author} {\bibfnamefont {D.}~\bibnamefont
  {Jackson Kimball}}, \bibinfo {author} {\bibfnamefont {T.}~\bibnamefont
  {Kornack}}, \bibinfo {author} {\bibfnamefont {V.}~\bibnamefont {Lebedev}},
  \bibinfo {author} {\bibfnamefont {W.}~\bibnamefont {Li}}, \bibinfo {author}
  {\bibfnamefont {H.}~\bibnamefont {Masia-Roig}}, \bibinfo {author}
  {\bibfnamefont {S.}~\bibnamefont {Nix}}, \bibinfo {author} {\bibfnamefont
  {M.}~\bibnamefont {Padniuk}}, \bibinfo {author} {\bibfnamefont
  {C.}~\bibnamefont {Palm}}, \bibinfo {author} {\bibfnamefont {C.}~\bibnamefont
  {Pankow}}, \bibinfo {author} {\bibfnamefont {A.}~\bibnamefont {Penaflor}},
  \bibinfo {author} {\bibfnamefont {X.}~\bibnamefont {Peng}}, \bibinfo {author}
  {\bibfnamefont {S.}~\bibnamefont {Pustelny}}, \bibinfo {author}
  {\bibfnamefont {T.}~\bibnamefont {Scholtes}}, \bibinfo {author}
  {\bibfnamefont {J.}~\bibnamefont {Smiga}}, \bibinfo {author} {\bibfnamefont
  {J.}~\bibnamefont {Stalnaker}}, \bibinfo {author} {\bibfnamefont
  {A.}~\bibnamefont {Weis}}, \bibinfo {author} {\bibfnamefont {A.}~\bibnamefont
  {Wickenbrock}}, \ and\ \bibinfo {author} {\bibfnamefont {D.}~\bibnamefont
  {Wurm}},\ }\href {\doibase https://doi.org/10.1016/j.dark.2018.10.002}
  {\bibfield  {journal} {\bibinfo  {journal} {Physics of the Dark Universe}\
  }\textbf {\bibinfo {volume} {22}},\ \bibinfo {pages} {162 } (\bibinfo {year}
  {2018})}\BibitemShut {NoStop}%
\end{thebibliography}
\end{document}